\documentclass{article}
\usepackage{arxiv}
\usepackage[utf8]{inputenc} % allow utf-8 input
\usepackage[T1]{fontenc}    % use 8-bit T1 fonts
\usepackage[colorlinks=true, allcolors=blue]{hyperref}       % hyperlinks
\usepackage{url}            % simple URL typesetting
\usepackage{booktabs}       % professional-quality tables
\usepackage{amsfonts}       % blackboard math symbols
\usepackage{nicefrac}       % compact symbols for 1/2, etc.
\usepackage{microtype}      % microtypography
\usepackage{lipsum}		% Can be removed after putting your text content
\usepackage{graphicx}
\usepackage{natbib}
\usepackage{doi}
\usepackage{adjustbox}
\usepackage[version=4]{mhchem}

\title{Detection of simplest amino acid {\it glycine\/} in the atmosphere of Venus}

\author{ {\color{red}Arijit  Manna}\\%\thanks{Use footnote for providing further
	Midnapore City College\\
	Kuturia, Bhadutala, Paschim Medinipur, \\West Bengal, India 721129 \\
	\texttt{Mannaarijit@hotmail.com} \\
	\And
	{\color{red}Sabyasachi Pal} \\
	Indian Centre for Space Physics\\43 Chalantika, Garia Station Road, \\Kolkata, India 700084\\
	\texttt{sabya.pal@gmail.com} \\
	\And
	{\color{red}Mangal Hazra} \\
	Midnapore City College\\
	Kuturia, Bhadutala, Paschim Medinipur, \\West Bengal, India 721129 \\
	\texttt{mangalkrhazra@gmail.com} \\
}

\renewcommand{\shorttitle}{\ce{NH2CH2COOH} in the atmosphere of Venus}

\begin{document}
\maketitle

\begin{abstract}
	Amino acids are considered to be prime ingredients in chemistry, leading to life. Glycine is the simplest amino acid and most commonly found in animal proteins. It is a glucogenic and non-essential amino acid that is produced naturally by living bodies and plays a key role in the creation of several other important bio-compounds and proteins. We report the first spectroscopic detection of the rotational absorption lines of the simplest amino acid glycine (\ce{NH2CH2COOH}) with confirmer I and II in the atmosphere of Venus using the archival data from the Atacama Large Millimeter/Submillimeter Array (ALMA). We detect the eleven rotational absorption lines of \ce{NH2CH2COOH} between the frequency range of $\nu$ = 245--262 GHz with $\geq$3$\sigma$ statistical significance. We calculate the total column density of glycine in the atmosphere of Venus is $N$(\ce{NH2CH2COOH}) $\sim$ 5$\times$10$^{14}$ cm$^{-2}$. Using the column density information of glycine, we calculate the abundance of glycine is $\sim$ 1.6$\times$10$^{-9}$ in the atmosphere of Venus. The detection of glycine in the atmosphere of Venus might be one of the keys to an understanding of the formation mechanisms of prebiotic molecules in the atmosphere of Venus. The detection of glycine indicates that the upper atmosphere of Venus may be going through nearly the same biological method as Earth billions of years ago.
\end{abstract}

% keywords can be removed

\section{Introduction}
\label{sec:intro} 
The averaged surface temperature of Venus is highest among solar system bodies ($\sim$740 K) \citep{Ma73} but in the middle and upper atmosphere, the temperature drops making a relatively hospitable environment for life. The temperature of clouds at the height of 50 km is 300--350 K with pressure around 1 bar, which is comparable to the ground temperature of Earth \citep{Co99}. So, Venus has been regarded as a possible sustainer of life for a long time \citep{Sa67, Gr07}. The atmosphere of Venus is extremely dense and mostly consists of the high amount of CO$_2$ (96.5 per cent), N$_2$ (3.5 per cent), and trace gases \citep{Ber07}. The mesosphere of Venus (60--120 km altitude) possesses complex cycles of dynamic processes and photochemistry that are still poorly understood. In the atmosphere of Venus, the complex compound \ce{NH4Cl} is produced at a height 30 km above the planet's surface level. The other ammonium compounds \ce{NH2COONH4} and \ce{NH4HCO3} are distribute up to an altitude of $\sim$ 60 km \citep{Ot74}. Earlier, \citet{Ot74} had shown in the laboratory that organic molecules, including glycine, can be produced in a Venus-like atmosphere with a gas mixture of N$_2$, NH$_3$, H$_2$O, O$_2$ and CO$_2$ in presence of a 60 kV spark. 

Recently, phosphine (PH$_{3}$) was claimed to be detected from the Venusian atmosphere at $\nu$ = 267 GHz using the Atacama Large Millimeter/Submillimeter Array (ALMA) \citep{Gre20}. The formation mechanism of \ce{PH3} in the Venusian atmosphere can not be explained by the conventional processes known to us and may be due to unknown geochemistry, photo-chemistry, or even aerial microbial life present in the upper Venusian atmosphere as Earth phosphine is mostly related with biological sources \citep{Ba20}. The detection of the rotational absorption line of phosphine in the atmosphere of Venus was questioned in many works of literature \citep{Sne20, Tho20}. After the controversy of the existence of \ce{PH3} in cloud decks of Venus, \citet{Enc20} presents the abundance of \ce{PH3} rotational line in the top cloud of the atmosphere of venus between 954 and 956 cm$^{-1}$ in infrared wavelength using TEXES. 

Glycine is the simplest amino acid and is regarded as the basic building block leading to life. On Earth, it is most commonly found in animal proteins. Glycine is a non-essential amino acid and is naturally formed by the living body on Earth and plays an important role in the development of many other important bio-compounds and proteins. Detection of biomolecules, especially amino acids, outside Earth gives important clues in the understanding of the formation and evolution of life outside our planet. It is generally believed that the formation of glycine may have occurred in the interstellar medium (ISM) as amino acids, including glycine, are detected in meteorites \citep{Pi91}. But, after a tentative detection of glycine in interstellar medium \citep{Ku03}, a rigorous study verifies that there is no proof of the presence of glycine in interstellar medium \citep{Sn05}.

In this letter, we present the spectroscopic detections of the simplest amino acid glycine with confirmer I and II absorption lines in the atmosphere of Venus using ALMA with 7m array ACA (Atacama Compact Array) data. In Sect.~\ref{sec:reduction}, we briefly describe the observations and data reductions using ALMA data. The spectroscopic detection of eleven rotational lines of \ce{NH2CH2COOH} with different transitions are presented in Sect.~\ref{sec:result} and discussion of the importance of glycine in the atmosphere of Venus is presented in Sect.~\ref{sec:discussion}.

\begin{figure*}
	\centering
	\includegraphics[width=0.49\textwidth]{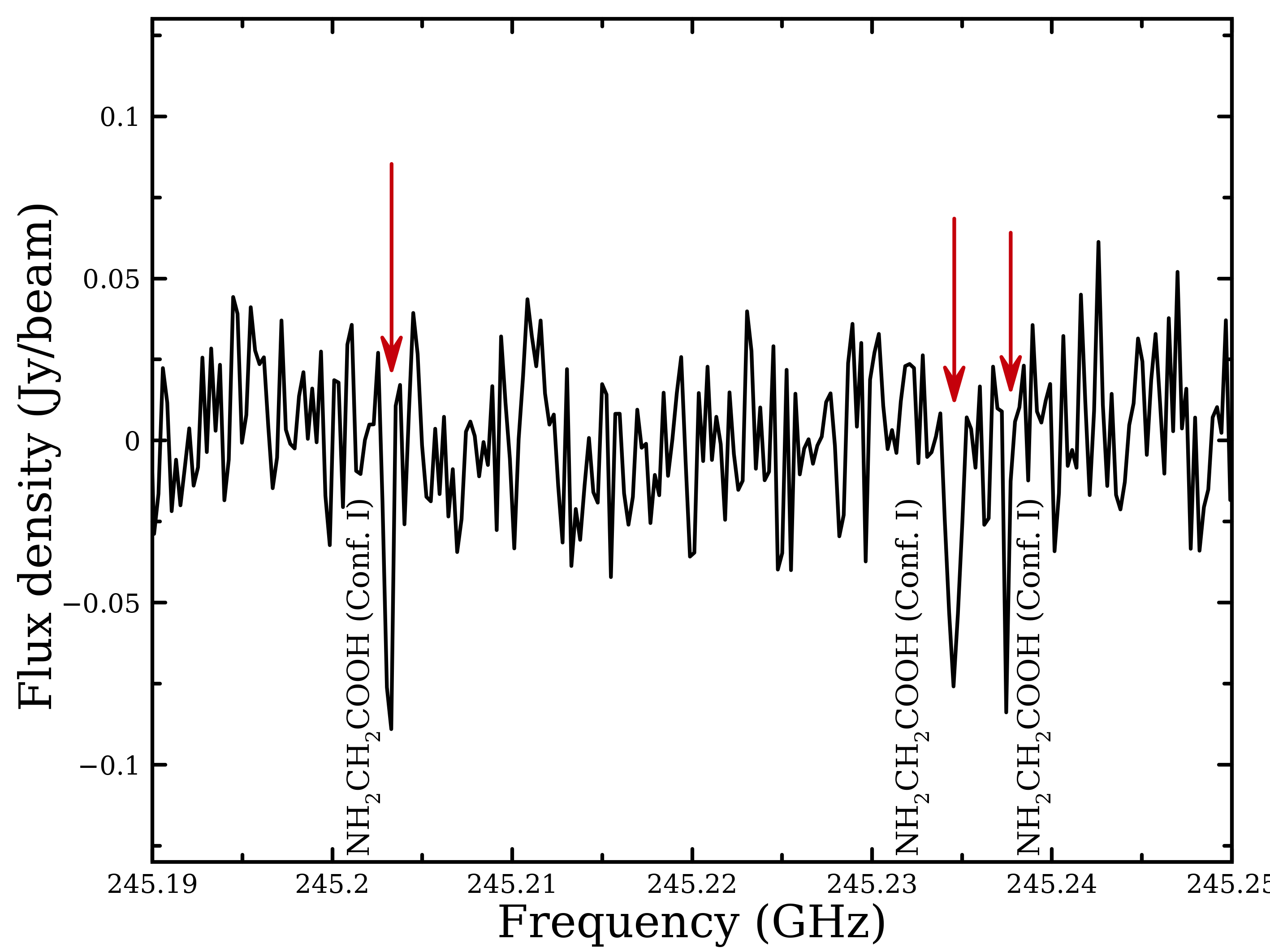}\includegraphics[width=0.49\textwidth]{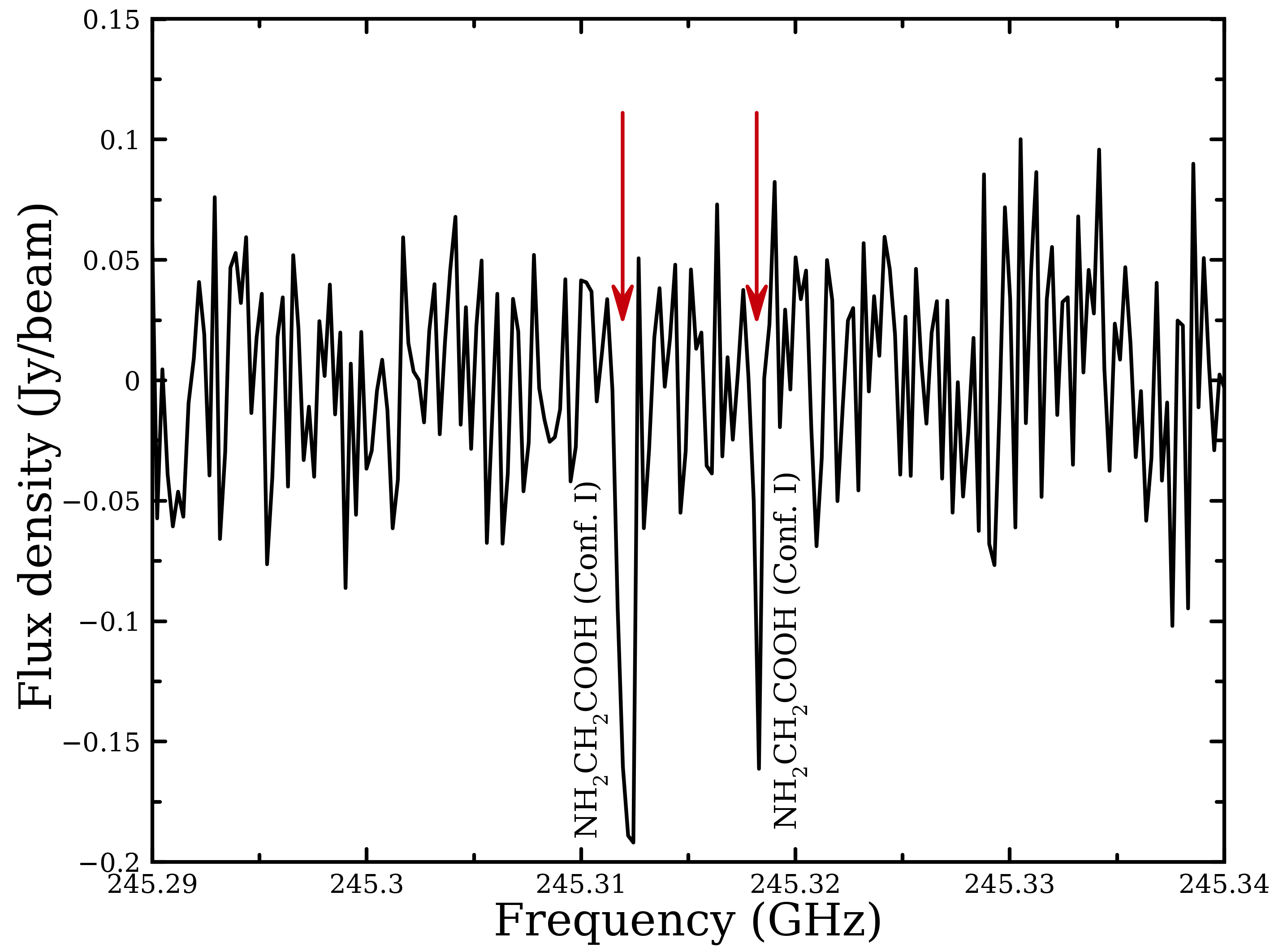}
	\includegraphics[width=0.49\textwidth]{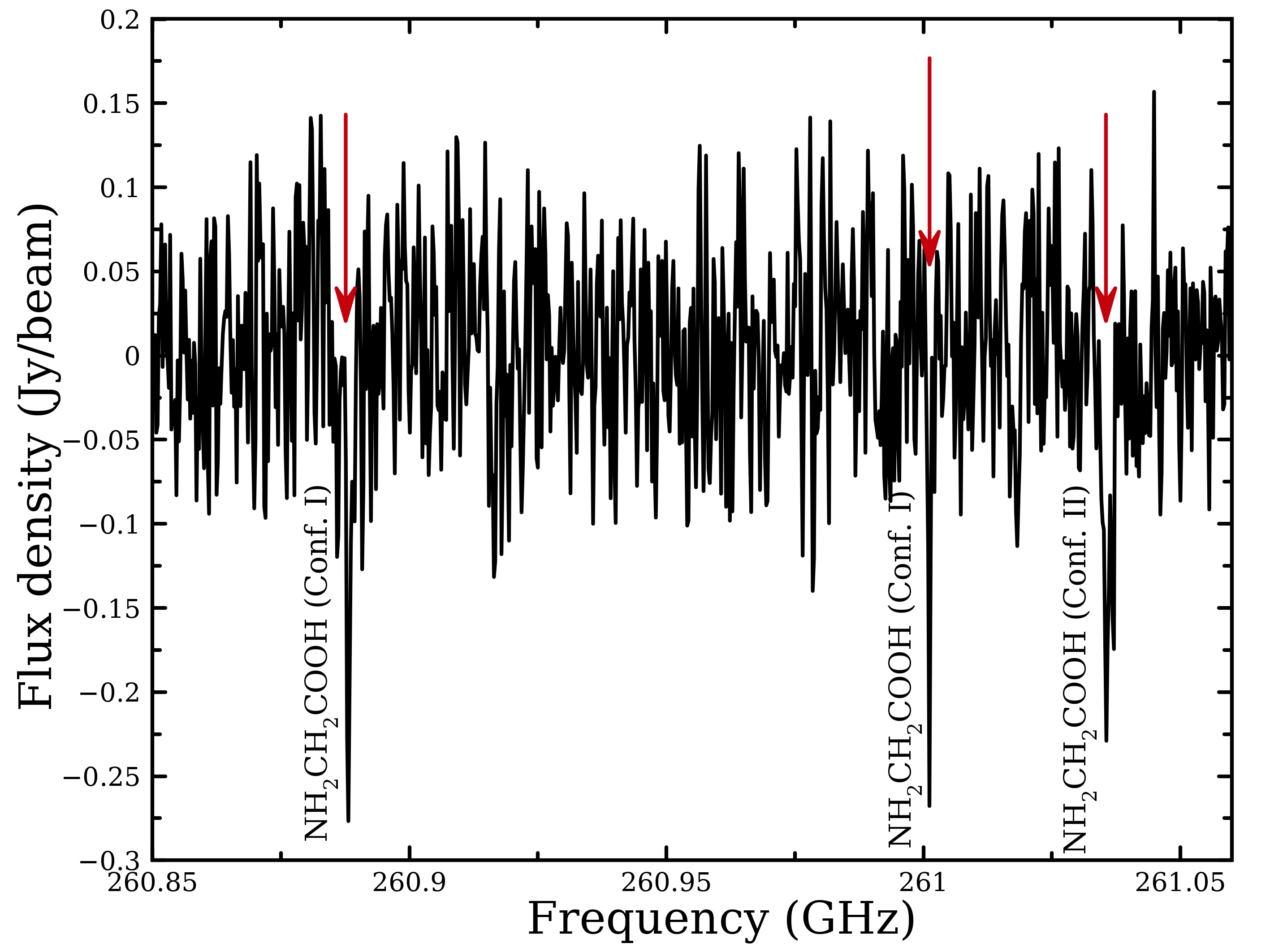}\includegraphics[width=0.49\textwidth]{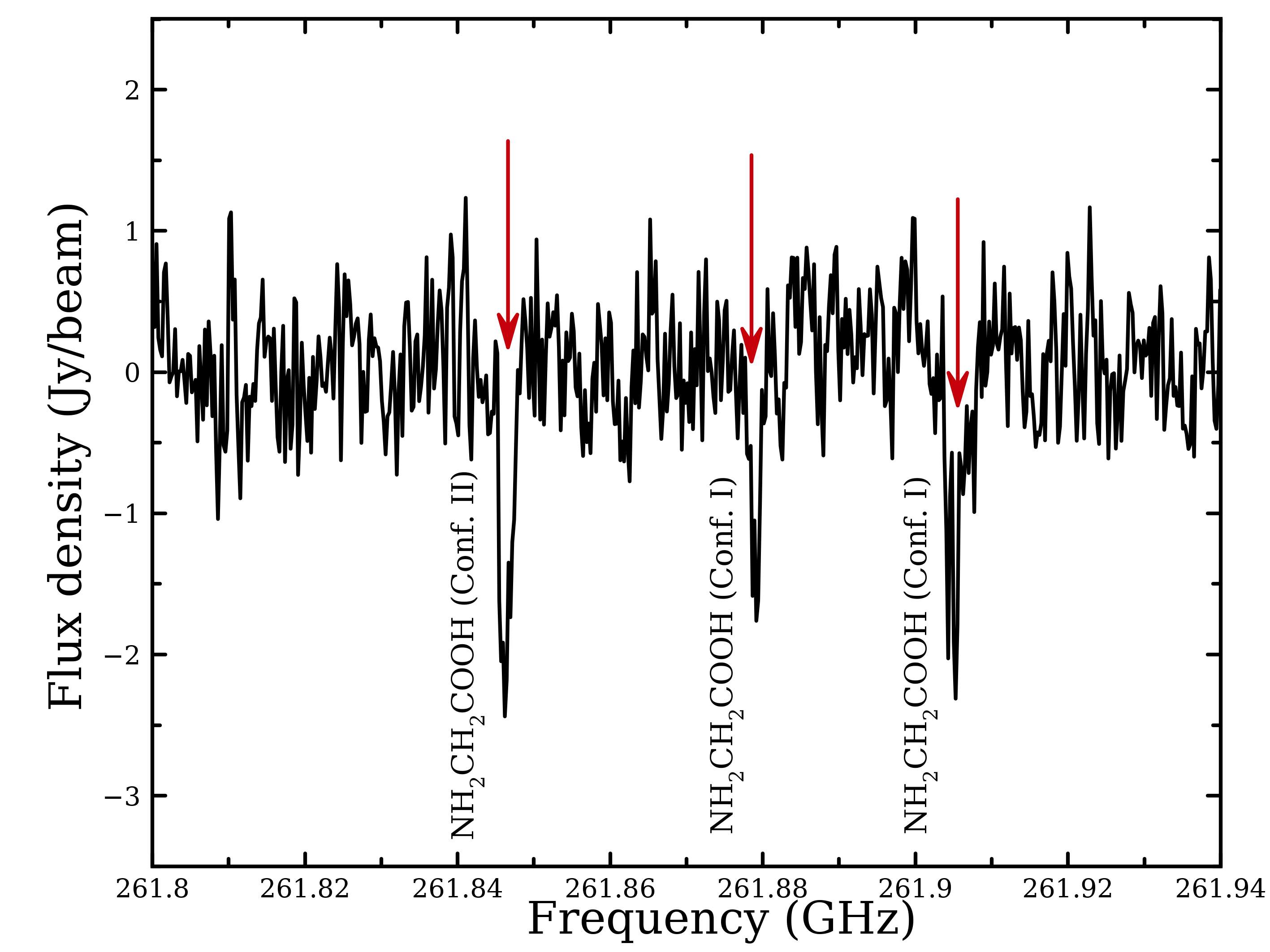}
	\caption{Molecular rotational absorption spectrum of the \ce{NH2CH2COOH} between the frequency range of $\nu$ = 245--262 GHz with diferent transitions using ALMA band 6 observation. The spectrum was made by integrating the reduced ALMA data cubes from the centre of Venus (12104 km at distance of Venus). The red arrow marks indicated the lines of glycine.}
	\label{fig:glycineimage}
\end{figure*}
%*******************************************************************************************************%

\section{Observation and data reduction}
\label{sec:reduction}

We used interferometric millimeter-wavelength data from the Atacama Large Millimeter/submillimeter Array (ALMA)\footnote{\href{https://almascience.nrao.edu/}{https://almascience.nrao.edu/}} with 7m array (ACA) data to study the molecular rotational lines in the atmosphere of Venus. The 40-minute observation took place on 8th January 2019 beginning at 11h05m UTC. During the observation, the angular diameter of Venus was 24.14$^{\prime\prime}$ with an illumination factor of 88.41 per cent and surface brightness 1.4 mag arcsec$^{-2}$. The correlator was configured so that eleven spectral windows were used inside ALMA Band 6 with a frequency range of 245.19--261.96 GHz. The spectral resolution of science spectral windows 16--26 is 242.31 kHz but for the case of spectral windows 28--36 the spectral resolution is 282.23 kHz. The XX and YY type signal substances were tested via an integration time of 390.97 seconds. A total of 11 antennas were accessible during observation in 7m ACA data with a minimum frequency resolution of 242.2826 kHz. The telescope was configured to track the NASA Horizon ephemeris position of Venus with real-time updating of the coordinates of the phase center.

The source J1256--0547 was used as a bandpass calibrator while J1512--0905 was used as a phase calibrator. Calibration and flagging of bad data were done by using the software package Common Astronomy Software Application ({\tt CASA 5.1.1-5})\footnote{\href{https://casa.nrao.edu/}{https://casa.nrao.edu/}}. The measured continuum flux density was set using the Butler-JPL-Horizons 2012 standard and the NASA ephemeris \citep{But12}. We used the task {\tt MSTRANSFORM} to split the target data after calibration and used {\tt UVCONTSUB} for the continuum subtraction of Venus. The emission from line-free channels in the calibrated visibility spectral tables was averaged to produce continuum visibilities for an image of the continuum emission from Venus. The continuum emission was subtracted from the visibility spectra by fitting the line-free channels with a polynomial function. The imaging was performed using the {\tt TCLEAN} task in CASA with several numbers of iterations and self-calibration. Hogbom algorithm was used to perform deconvolution of the point-spread function (PSF) for each spectral channel, with a threshold flux level of twice the expected RMS noise and natural visibility weighting. The primary beam of the resultant image of Venus after calibration is 7.308$^{\prime\prime}$$\times$3.454$^{\prime\prime}$ with position angle 78.835$^{\circ}$. The resultant spectrum of Venus was adjusted with NASA JPL Horizons Topocentric radial velocity activity with Doppler-shift to Venus rest frame.
\begin{table}
	\caption{Molecular properties of detected \ce{NH2CH2COOH} ratational absorption lines in the atmosphere of Venus. 1$\sigma$ errors and line fluxes were obtained from the continuum-subtracted ALMA spectra.}
	\centering
	%	\scriptsize
	\begin{tabular}{|c|c|c|c|c|c|c|c|c|c|} 		
		\hline
		Line & Frequency& Transition & E$_{l}$\\
		&(GHz) 	&    (J)       & (k) \\	
		\hline		
		\ce{NH2CH2COOH, Conf I}&245.203&36(7,30)-35(7,29)&237.306\\
		\ce{NH2CH2COOH, Conf I}&245.234&37(5,32)-36(6,31)&242.635\\
		\ce{NH2CH2COOH, Conf I}&245.237&65(20,45)-65(19,46)&842.237\\
		\ce{NH2CH2COOH, Conf I}&245.311&62(9,53)--62(8,54)&462.817\\
		\ce{NH2CH2COOH, Conf I}&245.318&36(32,4)-35(32,3)&379.660\\
		\ce{NH2CH2COOH, Conf I}&260.888&38(19,20)-37(19,19)&243.717\\
		\ce{NH2CH2COOH, Conf I}&261.000&76(22,54)-76(21,55)&776.542\\
		\ce{NH2CH2COOH, Conf II}&261.034&73(14,59)-73(14,60)&695.319\\
		\ce{NH2CH2COOH, Conf II}&261.845&63(22,41)-63(21,42)&580.347\\
		\ce{NH2CH2COOH, Conf I}&261.879&13(13,0)-12(12,1)&51.033\\
		\ce{NH2CH2COOH, Conf I}&261.905&24(8,17)-23(7,16)&74.965\\
		\hline
	\end{tabular}
	\label{tab:position}
\end{table}

\section{RESULT}
\label{sec:result}
\subsection{Rotational absorption lines of \ce{NH2CH2COOH} in the atmosphere of Venus} 
We used interferometric millimeter observation of Venus using the ALMA. Fig.~\ref{fig:glycineimage} shows the rotational absorption spectrum of \ce{NH2CH2COOH} in the Venusian atmosphere with confirmer I and II between the frequency range of $\nu$ = 245--262 GHz. The molecular properties of detected rotational lines of glycine including different transitions in the atmosphere of Venus are shown in Table.~\ref{tab:position}.

In the Venusian atmosphere, the spectrum was made by integrating the reduced ALMA data cubes from the center of Venus (12104 km at Venus's distance). The synthesized primary beam of the resultant image of Venus after calibration is 7.308$^{\prime\prime}$$\times$3.454$^{\prime\prime}$. The rotational spectral peaks of glycine were allocated to frequencies obtained from the Cologne Database for Molecular Spectroscopy (CDMS)\footnote{\href{https://cdms.astro.uni-koeln.de/cgi-bin/cdmssearch}{https://cdms.astro.uni-koeln.de/cgi-bin/cdmssearch}} \citep{Mu01}. The rotational absorption lines are easily detected with $\geq$3$\sigma$ statistical significance. Earlier, the amino acid glycine was detected in comet 81P/Wild 2 \citep{El09} and 67P/Churyumov-Gerasimenko \citep{Alt16} but this is the first time the presence of an amino acid is reported on a planet or moon. 

Glycine is detected strongly near the equator and mid-latitude in Venus. The distribution of glycine is stronger in mid-latitude (22.5$^\circ$--67.5$^\circ$). Near the pole, there is no evidence of the presence of glycine (<3$\sigma$). Recently, the presence of PH$_3$ in Venus was also found to be stronger near mid-latitude and it was not detected by ALMA beyond 60$^\circ$ latitude \citep{Gre20}. The mid-latitude Hadley circulation may give the most stable life-supporting condition with circulation times of 70--90 days being sufficient for (Earth-like) microbial life reproduction \citep{Gr07, Gre20}. At a height of 65--70 km, the zonal wind blows at a near-constant velocity $\sim$100 m s$^{-1}$ between latitude range 50$^\circ$N to 50$^\circ$S and then air speed gradually decreases towards the pole \citep{Sa17}. The latitude-dependent distribution of glycine roughly matches (within $\sim$10$^\circ$) with the detection limit of recently detected phosphine \citep{Gre20} and with the proposed upper Hadley-cell boundary \citep{Sa17} where gas circulates between upper and lower altitudes.

\subsection{Abundance of glycine in the Venusian atmosphere}
At first, we assume that the rotational molecular lines of glycine are optically thin and Local Tharmal Eqilibrium (LTE) populated as well as the upper limit of $T_{mb}$ is directly proportional to the column density of glycine using the following relation \citep{Co96}\\
%\begin{multline}
%\scriptsize
%	\begin{adjustbox}{width=0.5\textwidth}
$N_{(glycine)} = \frac{3Z}{8 \pi^3}
\frac{T_{mb}\delta v}{\mu^2 S_{ul}}\frac{k}{\nu}\exp\biggr(\frac{E_u}{kT}\biggr)
\times\biggr[1-\frac {\exp(\frac{h\nu}{kT}-1)}{\exp(\frac{h\nu}{kT_{bg}}-1)}\biggr]^{-1}$-----(1)\\
%\end{adjustbox}
%\label{eq:ntotfinal}
%\end{multline} 
where $S_{ul}$ is the line strength of detected absorption lines of glycine, $E_{u}$ is the upper energy in K, $T_{bg}$ is the background temperature, $T$ is the gas temperature, $\mu$ denoted as the dipole moment, and $Z$ denoted as the partition function. We used the CDMS to determine the significance value of statistical parameters like $E_{u}$, $S_{ul}$, and partition function $Z$. On the dependence of gas temperature, the column density (average on the beam) of the glycine is $N$(\ce{NH2CH2COOH})$\sim5\times$10$^{14}$ cm$^{-2}$ with temperature $T$ = 75 K. In a high energy, the dipole moment of the confirmer II line is $\sim$ 5 times higher than that of the confirmer I line. The abundance of glycine in the atmosphere of Venus is $\sim$ 1.6$\times 10^{-9}$ with respect to \ce{CO2} where column density of \ce{CO2} is $N$(\ce{CO2}) $\sim$ 3$\times 10^{23}$ cm$^{-2}$. 

\subsection{Microbial life in the cloud of Venus}
In 1997, \citet{Sa67} suggested a possibility in which multicellular organisms with a diameter of up to 4 cm could live in the upper atmosphere of Venus. %\citet{ks13} have been photographed the multicellular organisms at ground level. 
The high abundance of deuterium and trace quantities of water can be found in the cloud of Venus, which may support the creation of life \citep{Don93}. The lower cloud layer of Venus (47.5--50.5 km) provides ideal conditions for microbial life, including reasonable temperatures and pressures (60$^{\circ}$C and 1 atm), and the presence of micron-sized sulfuric acid aerosols, which may support sulfate-reducing microbes \citep{lim18}. \citet{sch04} give a theory that concludes that Venusian clouds, which are 48 to 65 kilometers above the surface, may be home to sulfur-metabolizing aeroplankton and acidophilic microbes that use sulfur allotropes as photo-protective pigments to allow photosynthesis.
The wind speeds on Venus can approach 355 km h$^{-1}$ in the upper cloud layers and may rise to more than 700 km h$^{-1}$ in the lower cloud layers \citep{Bl86}. The growth of phytoplankton and diatoms can be inhibited by increasing turbulence. Our detection of glycine in the atmosphere of Venus gives hints for the formation as well as the existence of microbial life.

\section{Conclusion and Discussion}
\label{sec:discussion}
In this letter, we report the first spectroscopic detection of the simplest amino acid glycine in the atmosphere of Venus with confirmer I and II using ALMA band 6 observation. We present the eleven rotational absorption lines of glycine between the frequency range of $\nu$ = 245--262 GHz with $\geq$3$\sigma$ statistical significance. The statistical column density of glycine is $N$(\ce{NH$_{2}$CH$_{2}$COOH}) $\sim 5\times$10$^{14}$ cm$^{-2}$ and abundance of glycine with respect to \ce{CO2} is $\sim$ 1.6$\times 10^{-9}$. 

Miller and Urey's experiment in 1953 \citep{Mill53} simulated the primordial Earth conditions and tested the chemical origin of life in the Earth atmosphere. The experiment used water (H$_{2}$O), methane (CH$_{4}$), ammonia (NH$_{3}$), and hydrogen (H$_{2}$) and produced many organic compounds including glycine and different types of other amino acids. A recent experiment with the preserved laboratory materials of Miller synthesized more than 40 different amino acids and amines which shows possibilities of formation of biological compounds under different cosmochemical conditions \citep{Ba13}. Miller and Urey's experiment found glycolic acid (CH$_{2}$OHCOOH) which produces the simple amino acid glycine when it reacts with ammonia (NH$_{3}$) (CH$_{2}$OHCOOH + NH$_{3}$ $\longrightarrow$ CH$_{2}$NH$_{2}$COOH + H$_{2}$O). Since HDO \citep{Don93}, CH$_{4}$ \citep{Don93},
NH$_{3}$ \citep{Goe74}, and CO$_{2}$ \citep{Sne14} are already detected in the Venus atmosphere, glycine may be formed following the route of Miller and Urey experiment. Alternatively, glycine may have been also formed by a reaction between NH$_{3}$, CH$_{2}$ and CO$_{2}$ all of which are already present in Venus atmosphere (NH$_{3}$+CH$_{2}$+CO$_{2}$ $\longrightarrow$ CH$_{2}$NH$_{2}$COOH) \citep{Mea04}. Recently, amino acid decomposition has been reported in high-pressure and high-temperature water \citep{Sat04}. These decomposition reactions in condensed systems seem to be irreversible, in the gas phase, there may be reverse reaction routes producing amino acids.

In astrophysics, chemical physics, and biophysics, synthetic reaction routes of the simplest amino acid glycine, from simple molecules have great significance with chemical evolution and the origin of life \citep{Gu10}. The detection of glycine in the atmosphere of Venus may indicate the existence of an early form of life in the atmosphere of the solar planet because amino acid is a building block of protein \citep{Nor17}. Venus may be going through the primary stage of biological evolution. It should be noted that the detection of glycine in the Venusian atmosphere is a hint of the existence of life but not robust evidence. Glycine may also be produced in the atmosphere of Venus by other photochemical or geochemical means, not common on Earth. A Venus mission (such as proposed by \citet{Hei20}) with direct sampling from Venusian surface and cloud may confirm the source of glycine in the planet.

\subsection*{Acknowledgements}
This paper makes use of the following ALMA data: ADS/JAO.ALMA\#2018.1.00879.S. ALMA is a partnership of ESO (representing its member states), NSF (USA), and NINS (Japan), together with NRC (Canada), MOST and ASIAA (Taiwan), and KASI (Republic of Korea), in cooperation with the of Chile. The Joint ALMA Observatory is operated by ESO, AUI/NRAO, and NAOJ.

\section*{Data Availability Statement}
The data that support the plots within this paper and other findings of this study are available from the corresponding author upon reasonable request. The raw ALMA data are publicly available at \href{https://almascience.nao.ac.jp/asax/}{https://almascience.nao.ac.jp/asax/} (project id : 2018.1.00879.S).


\begin{thebibliography}{}
	\bibitem[{Altwegg et al.}(2016)]{Alt16} Altwegg K. et al., 2016, Science Advances, 2.
	
	\bibitem[{Bada}(2013)]{Ba13} Bada J. L., 2013, Chem. Soc. Rev, 42, 2186.
	
	\bibitem[{Bains et al.}(2020)]{Ba20} Bains W. et al., 2020, arXiv:2009.06499.
	
	\bibitem[{Bertaux et al.}(2007)]{Ber07} Bertaux J. et al., 2007, Nature, 5, 646.
	
	\bibitem[{Blamont et al.}(1986)]{Bl86}Blamont J. E. et al., 1986, Science, 231, 1422
	
	\bibitem[{Butler}(2012)]{But12} Butler B., 2012, ALMA Memo 594, NRAO.
	
	
	
	\bibitem[{Cockell}(1999)]{Co99} Cockell C.~S., 1999, Planet. Space Sci., 47, 1487.
	
	\bibitem[{Combes et al.}(1996)]{Co96}Combes F., Nguyen R., Wlodarczak G., 1996, A\&A, 308, 618.
	
	\bibitem[{Donahue \& Hodges}(1993)]{Don93} Donahue T., Hodges R. R., 1993, Geophys. Res. Lett., 20, GL00513.
	
	
	\bibitem[{Elsila et al.}(2009)]{El09} Elsila J. E., Glavin D. P., Dworkin J. P., 2009, Meteoritics \& Planetary Science, 44, 1323.
	
	\bibitem[{Encrenaz et al.}(2020)]{Enc20} Encrenaz T. et al., 2020, arXiv:2010.07817v1.
	
	\bibitem[{Greaves et al.}(2020)]{Gre20} Greaves J.~S. et al., 2020, Nature Astronomy.
	
	\bibitem[{Griffith et al.}(2006)]{Gr06} Griffith C. A. et al., 2006, Science, 313, 1620.
	
	\bibitem[{Grinspoon \& Bullock}(2007)]{Gr07} Grinspoon D. H., Bullock M. A., 2007, AGU, 191.
	
	\bibitem[{Goettel \& Lewis}(1974)]{Goe74} Goettel K., Lewis J., 1974, Journal of The Atmospheric Sciences, 828.
	
	\bibitem[{Guti{\'e}rrez-Preciado et al.}(2010)]{Gu10} Guti{\'e}rrez-Preciado A., Romero H., Peimbert M., 2010, Nature Education, 3, 29.
	
	\bibitem[{Hein et al.}(2020)]{Hei20} Hein A. M. et al., 2020, arXiv:2009.11826v2.
	
	\bibitem[{Hueso \& S{\'a}nchez-Lavega}(2006)]{Hu06} Hueso R., S{\'a}nchez-Lavega A., 2006, Nature., 442, 428.
	
	%	\bibitem[{Ksanfomality}(2013)]{ks13} Ksanfomality L.W., 2013, Dokl. Phys.58, 204
	
	\bibitem[{Kuan et al.}(2003)]{Ku03} Kuan Y. J., Charnley S. B., Huang H. C., Tseng W. L., Kisiel Z. 2003, ApJ, 593, 848.
	
	\bibitem[{Lopes et al.}(2019)]{Lo19} Lopes R. M. C. et al., 2019, Space Sci. Rev., 215, 33.
	
	\bibitem[{Limaye et al.}(2018)]{lim18}Limaye S. S. et al., 2018, Epub, PMC6150942.
	
	\bibitem[{Marov et al.}(1973)]{Ma73} Marov M.~Y. et al., 1973, Icarus., 20, 407.
	
	\bibitem[{Maeda \& Ohno}(2004)]{Mea04} Maeda S., Ohno K., 2004, Chemical Physics Letters, 398, 240.
	
	\bibitem[{Miller}(1953)]{Mill53} Miller S. L., 1953, Science, 117, 528.
	
	\bibitem[{Moroz et al.}(1986)]{Mor86} Moroz V. I., Keating G. M., Kliore A., 1986, Committee on Space Research, 5, 303.
	
	\bibitem[{M{\"u}ller at al.}(2001)]{Mu01} M{\"u}ller H. S. P., Thorwirth S., Roth D. A., Winnewisser G., 2001, A\&A., 370, L49--L52.
	
	\bibitem[{Norio \& Shigenori}(2017)]{Nor17} Norio K., Shigenori M., 2017, Geoscience Frontiers, 9.
	
	\bibitem[{Otroshchenko \& Surkov}(1974)]{Ot74} Otroshchenko V. A., Surkov Yu. A., 1974, Origins Life Evol Biosphere, 5, 487--490.
	
	\bibitem[{Pizzarello et al.}(1991)]{Pi91} Pizzarello S., Krishnamurthy R. V., Epstein S., Cronin J. R., 1991, Geochimica Cosmochimica Acta., 55, 905.
	
	\bibitem[{Sagan \& Morowtz}(1967)]{Sa67} Sagan C., Morowtz H., 1967, Nature, 216, 1198.
	
	\bibitem[{S{\'a}nchez-Lavega et al.}(2017)]{Sa17} S{\'a}nchez-Lavega A., Lebonnois S., Imamura T., Read P., Luz D., 2017, Space Sci. Rev., 212, 1541.
	
	\bibitem[{Sato et al.}(2004)]{Sat04} Sato N., Quitain A. T., Kang K., Daimon H., Fujie K., 2004, Ind. Eng. Chem., 43, 3217.
	
	\bibitem[{Schulze-Makuch et al.}(2004)]{sch04}Schulze-Makuch D., Grinspoon D. H., Abbas O., Irwin L. N., Bullock M. A., 2004, Astrobiology, 4, 11.
	
	\bibitem[{Snels et al.}(2014)]{Sne14} Snels M., Stefani S., Grassi D., Piccioni G., Adriani A., 2014, Planet. Space Sci., 103, 347.
	
	\bibitem[{Snellen et al.}(2020)]{Sne20} Snellen I. A. G. et al., 2020, arXiv:2010.09761v1.
	
	\bibitem[{Snyder et al.}(2005)]{Sn05} Snyder L. E. et al., 2005, ApJ, 619, 914.
	
	\bibitem[{Thompson}(2020)]{Tho20} Thompson M. A., 2020, arXiv:2010.15188v1. 
	
	
\end{thebibliography}
\end{document}